\documentclass[sigconf]{acmart}
\usepackage{comment}
\usepackage{enumitem}
\usepackage{multirow}
\usepackage{flushend,}
\usepackage{ulem} 
\usepackage{graphicx}
\usepackage{subfig}
\usepackage{balance}

\makeatletter  
\newif\if@restonecol  
\makeatother

\usepackage[linesnumbered,ruled,vlined]{algorithm2e}
\usepackage{algpseudocode}  
\usepackage{amsmath}  

\AtBeginDocument{%
  \providecommand\BibTeX{{%
    \normalfont B\kern-0.5em{\scshape i\kern-0.25em b}\kern-0.8em\TeX}}}

\setcopyright{acmcopyright}
\copyrightyear{2018}
\acmYear{2018}
\acmDOI{10.1145/1122445.1122456}

\acmConference[Woodstock '18]{Woodstock '18: ACM Symposium on Neural
  Gaze Detection}{June 03--05, 2018}{Woodstock, NY}
\acmBooktitle{Woodstock '18: ACM Symposium on Neural Gaze Detection,
  June 03--05, 2018, Woodstock, NY}
\acmPrice{15.00}
\acmISBN{978-1-4503-XXXX-X/18/06}

\copyrightyear{2021}
\acmYear{2021}
\setcopyright{acmcopyright}\acmConference[WSDM '21]{Proceedings of the Fourteenth ACM International Conference on Web Search and Data Mining}{March 8--12, 2021}{Virtual Event, Israel}
\acmBooktitle{Proceedings of the Fourteenth ACM International Conference on Web Search and Data Mining (WSDM '21), March 8--12, 2021, Virtual Event, Israel}
\acmPrice{15.00}
\acmDOI{10.1145/3437963.3441811}
\acmISBN{978-1-4503-8297-7/21/03}

\begin{document}
\settopmatter{printacmref=false, printfolios=false}

\title{Sparse-Interest Network for Sequential Recommendation}

\author[Q. Tan, J. Zhang, J. Yao, N. Liu, J. Zhou, H. Yang, X. Hu]{
    Qiaoyu Tan$^{1}$, Jianwei Zhang$^{2}$, Jiangchao Yao$^{2}$, Ninghao Liu$^{1}$
}
\author{
    Jingren Zhou$^{2}$, Hongxia Yang$^{2}$, Xia Hu$^{1}$
}
\affiliation{
    $^1$ Department of Computer Science and Engineering, Texas A\&M University, TX, USA
}
\affiliation{
    $^2$ Alibaba Group
}

\email{
  {qytan,nhliu43,xiahu}@tamu.edu
}
\email{
  {zhangjianwei.zjw, jiangchao.yjc, jingren.zhou, yang.yhx}@alibaba-inc.com
}

\fancyhead{}

\begin{abstract}
Recent methods in sequential recommendation focus on learning an overall embedding vector from a user's behavior sequence for the next-item recommendation. However, from empirical analysis, we discovered that a user's behavior sequence often contains multiple conceptually distinct items, while a unified embedding vector is primarily affected by one's most recent frequent actions. Thus, it may fail to infer the next preferred item if conceptually similar items are not dominant in recent interactions.
To this end, an alternative solution is to represent each user with multiple embedding vectors encoding different aspects of the user's intentions. Nevertheless, recent work on multi-interest embedding usually considers a small number of concepts discovered via clustering, which may not be comparable to the large pool of item categories in real systems. It is a non-trivial task to effectively model a large number of diverse conceptual prototypes, as items are often not conceptually well clustered in fine granularity. Besides, an individual usually interacts with only a sparse set of concepts. 
In light of this, we propose a novel \textbf{S}parse \textbf{I}nterest \textbf{NE}twork (SINE) for sequential recommendation. Our sparse-interest module can adaptively infer a sparse set of concepts for each user from the large concept pool and output multiple embeddings accordingly. Given multiple interest embeddings, we develop an interest aggregation module to actively predict the user's current intention and then use it to explicitly model multiple interests for next-item prediction.
Empirical results on several public benchmark datasets and one large-scale industrial dataset demonstrate that SINE can achieve substantial improvement over state-of-the-art methods.

\end{abstract}

\ccsdesc[500]{Computer systems organization~Embedded systems}
\ccsdesc[300]{Computer systems organization~Redundancy}
\ccsdesc{Computer systems organization~Robotics}
\ccsdesc[100]{Networks~Network reliability}

\keywords{Recommender system, Sequential recommendation, Sparse-interest network, Multi-interest extraction}

\maketitle

{\fontsize{8pt}{8pt} \selectfont \textbf{ACM Reference Format:}\\
Qiaoyu Tan, Jianwei Zhang, Jiangchao Yao, Ninghao Liu, Jingren Zhou, Hongxia Yang, Xia Hu. 2021. Sparse-Interest Network for Sequential Recommendation. In \it Proceedings of the Fourteenth ACM International Conference on Web Search and Data Mining (WSDM ’21), March 8–12, 2021, Virtual Event, Israel. ACM, New York, NY, USA, 9 pages. \href{https://urldefense.com/v3/__https://doi.org/10.1145/3437963.3441811__;!!KwNVnqRv!QMsEAQ0ZwWc1HRCEEsWnlhXxKCkFuu52gYUDNzFzC7Zn5TnD28xpyI9FaZWGQ6g}{\url{https://doi.org/10.1145/3437963.3441811} }}

\begin{figure}[!t]
\includegraphics[width=8.8cm,height=3.8cm]{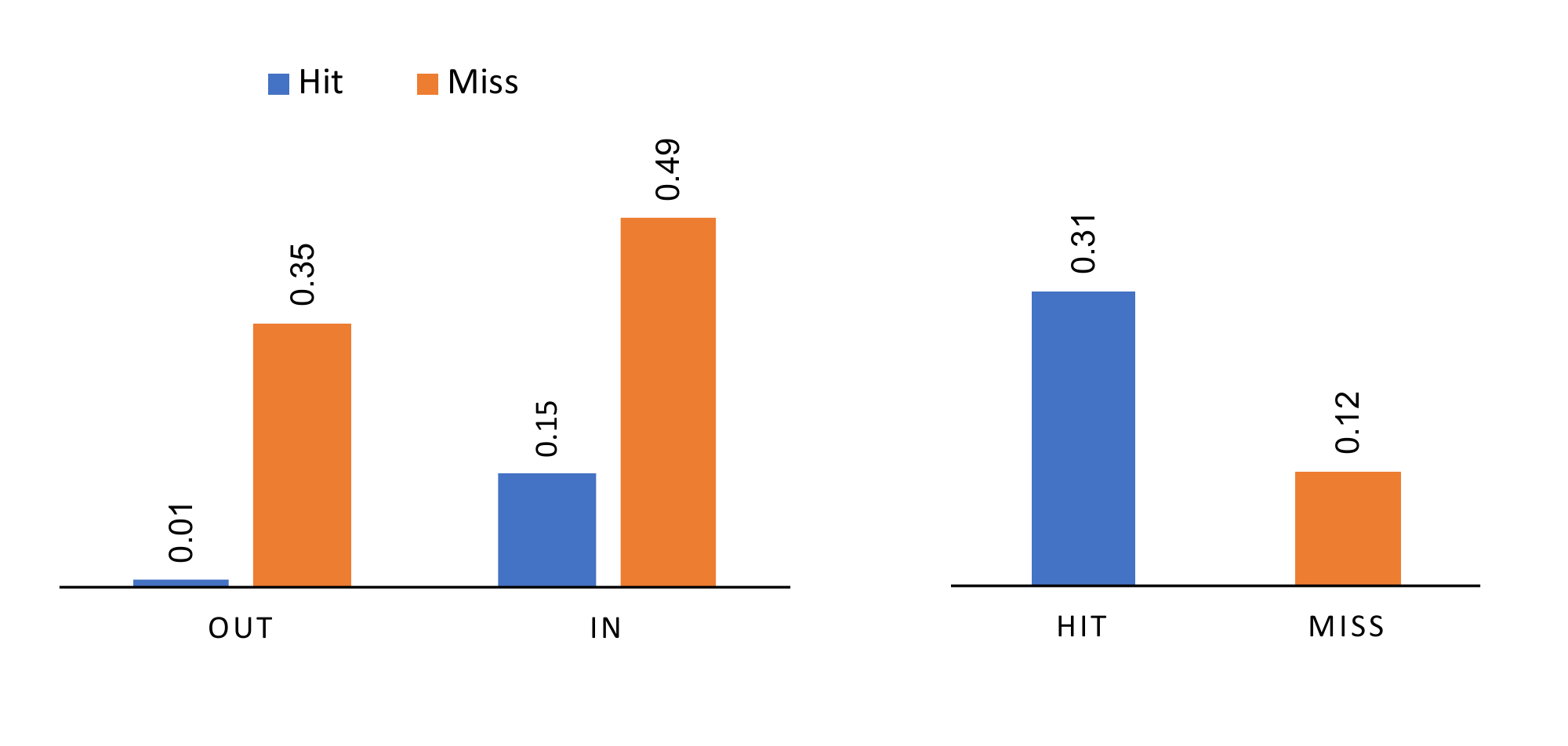}
\caption{Hit and Miss analysis in top@100 of single-embedding based SASRec~\cite{kang2018self} for next-item prediction on Taobao~\cite{zhu2018learning}. The left side shows the prediction results over "In" and "Out" settings. "In" means similar items belong to the same category of next predicted item are interacted in recent fifty behaviors, otherwise "Out". The right side shows the frequency of similar items in recent five behaviors. SASRec prefers to correctly predict the next-item if similar items are dominant in past interactions. }
\label{fig1}
\end{figure}
\section{Introduction}\label{sec:intro}
Recommender systems have been widely applied to many online services such as E-commerce, advertising, and social media to perform personalized information filtering~\cite{paul2016,hidasi2015session,ying2018graph,he2017neural}. At its core is to estimate how likely a user will interact with an item based on the past actions, e.g., purchases and clicks. Traditional recommendation methods adopt collaborative filtering approaches~\cite{sarwar2001item} to address the problem by assuming that behaviorally similar users would exhibit similar preferences on items. Recently, neural-based deep recommendation models have shown revolutionary performance in many recommendation scenarios, due to the powerful expressive ability of deep learning. For example, NCF~\cite{he2017neural} extends matrix factorization based models~\cite{sarwar2001item} by replacing the interaction function of inner product with nonlinear neural networks. PinSage~\cite{ying2018graph} is built on GraphSage~\cite{hamilton2017inductive}, and learns user and item embeddings by conducting convolutional operations on the user-item interaction graph. 
However, these methods ignore the sequential structure in user behaviors and thus fail to capture the correlations between adjacent behaviors.  

Some recent works formalize recommendation as a sequential problem. The principal idea behind this is to represent each user with an ordered sequence and assume its order matters. With a user's behavior history, the sequential recommendation approach first sorts the past behaviors to obtain the ordered sequence. After that, the sequence will be fed into different neural sequential modules (e.g., recurrent neural network~\cite{hidasi2015session}, convolutional network~\cite{tang2018personalized}, and Transformer~\cite{kang2018self}) to generate an overall user embedding vector, which is then used to predict the next interested item. Since the sequential recommendation approach reflects the real-world recommendation situation, it has 
attracted much attention in modern recommendation systems. 

Despite the recent advances, we argue that existing sequential recommendation models may be sub-optimal for next-item prediction due to the bottleneck of learning a single embedding from the user's behavior sequence. Each user in an E-commerce platform usually interacts with several types of items over time that are conceptually different. For example, we find that the number of categories of items that belong to different categories in a user's recent fifty behaviors is around 10 on Taobao dataset~\cite{zhu2018learning}. With multiple user's intentions~\footnote{Through out the paper, we interchangeably use intention and interest to indicate item cluster that consists of conceptually similar items. }, we also observe that, in Figure~\ref{fig1}, an overall user embedding vector learned from a behavior sequence is primarily affected by the recent most frequent actions. Thus, it may fail to extract related information for learning to predict the next item if its conceptually similar items are not dominant in recent interactions. Therefore, a promising alternative solution is to learn multiple embedding vectors from a user's behavior sequence, where each embedding vector encodes one aspect of the user's interests.

However, there are several challenges for effectively extracting multiple embedding vectors from the user's behavior sequence in industry-level data. First, items are often not conceptually well clustered in real systems. Although category information of items can be used as concepts, in many cases, such type of auxiliary information may not be available or reliable due to annotation noise in practice. The second challenge is to adaptively infer a sparse set of interested concepts for a user from the large concept pool. The inference procedure includes a selection operation, which is a discrete optimization problem and hard to train end-to-end. Third, given multiple interest embedding vectors, we need to determine which interest is likely to be activated for next-item predictions. During training, the next predicted item could be used as a label to activate the preferred intention, but the inference stage has no such label. The model has to predict a user's next intention adaptively. 

In this paper, we propose a novel Sparse-Interest NEtwork (SINE) for sequential recommendation to address these issues. SINE can learn a large pool of interest groups and capture multiple intentions of users in an end-to-end fashion. Figure~\ref{fig2} shows the overall structure of SINE. Our sparse interest extraction module adaptively infers the interacted interests of a user from a large pool of interest groups and outputs multiple interest embeddings. The aggregation module enables dynamically predicting the user's next intention, which helps to capture multi-interests for top-N item recommendation explicitly. We conduct experiments on several public benchmarks and an industrial dataset. Empirical results show that our framework outperforms state-of-the-art models and produces reasonable item clusters. To summarize, the main contributions of this paper are: 
\begin{itemize}
\item We propose a comprehensive framework that integrates large-scale item clustering and sparse-interest extraction jointly in a recommender system.   

\item We investigate an adaptive interest aggregation module to explicitly model users' multiple interests for top-N recommendation in the sequential recommendation scenario. 

\item Our model not only achieves state-of-the-art performance on several real-world challenging datasets, but also produces reasonable interest groups to assist multi-interest extraction.
\end{itemize}

\section{Related Work}\label{sec:relatework}
\begin{figure*}[h]
  \centering
  \includegraphics[width=15.0 cm,height=6.0 cm]{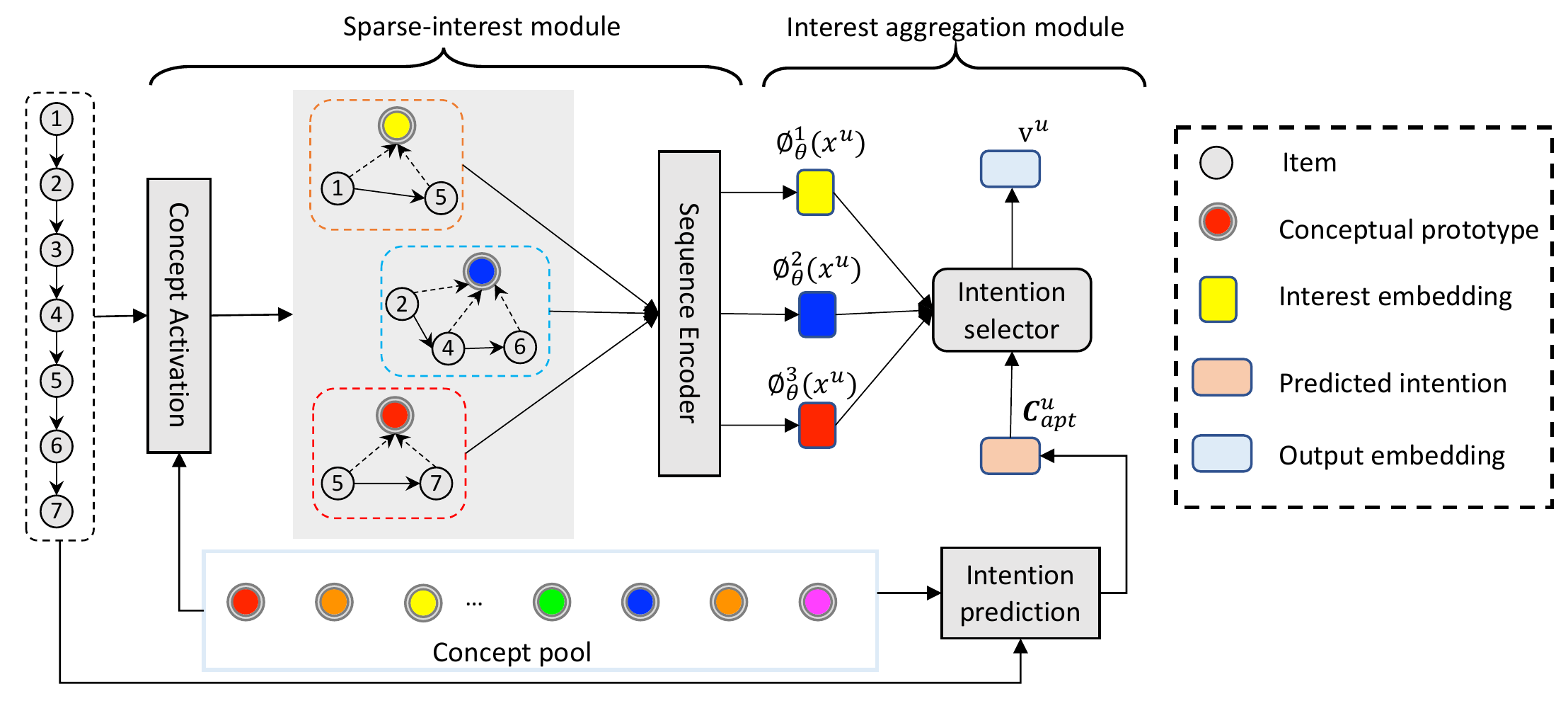}
  \caption{The architecture of SINE (better viewed in color). Given a user's behavior sequence as input, sparse-interest module aims to adaptively activate his/her interests from the large interest group pool as well as output multi-interest embeddings. Then, the interest aggregation module helps to select the most preferred interest for next-item recommendation by actively predicting user's next intention. SINE offers the ability to cluster items and infer user's sparse set of interests in an end-to-end fashion.}
  \label{figure1}
\end{figure*}

\subsection{General Recommendation}
In the conventional recommendation system, researchers focus on extracting users' general tastes from their historical behaviors. The typical examples include collaborative filtering~\cite{sarwar2001item,schafer2007collaborative}, matrix factorization techniques~\cite{koren2009matrix}, and factorization machines~\cite{rendle2010factorization}. The critical challenge of them lies in representing users and items with embedding vectors to compute their similarity. Matrix factorization (MF) methods seek to map users and items into joint latent space and estimate user-item interactions through the inner product between their embedding vectors. Factorization machines~\cite{rendle2010factorization} aim to model all interactions between variables using factorized parameters and can even estimate interactions when facing sparsity problems.   

Recently, inspired by the success of deep learning in computer vision and natural language processing~\cite{zhang2019deep}, much effort has been put into developing deep-learning-based recommender algorithms~\cite{he2017neural,guo2017deepfm,tan2020learning}. One line of work seeks to use neural networks to extract additional features for the content-aware recommendation~\cite{kim2016convolutional}. Another range of work targets to replace traditional MF. For example, NCF~\cite{he2017neural} uses multi-layer perceptions to replace the inner product operation in MF for interaction estimation, while AutoRec~\cite{sedhain2015autorec} adopts autoencoders to predict ratings. Moreover, several attempts also tried to apply graph neural networks~\cite{fan2019graph,jin2020multi,tan2019deep,yu2020graph} for recommendation~\cite{he2020lightgcn,ying2018graph}. 

\subsection{Sequential Recommendation}
Sequential recommendation has become the crucial problem of modern recommender systems, owing to its ability to capture the sequential patterns among successive items. One line of work attempts to model the item-to-item transition matrix based on the Markov Chain (MC). For instance, some works model the sequence using first-order Markov chain~\cite{rendle2010factorizing,cheng2013you}, which assumes that the next action only relies on the last behavior. To relax this limitation, there are also methods adopting high-order MCs that consider more previous items~\cite{he2016fusing,he2016vista,yan2019cosrec}. A representative work is Caser~\cite{tang2018personalized}, which treats use's behavior sequence as an "image" and adopts Convolutional Neural Network to extract user representation. 

Another line of works seeks to use a sequential neural module to process the user behavior sequence~\cite{kang2018self,hidasi2018recurrent,sun2019bert4rec,tan2021dman}. For example, 
GRU4Rec~\cite{hidasi2015session} first applies Gated Recurrent Units (GRU) to model the whole session for a more accurate recommendation. At the same time, SASRec~\cite{kang2018self} explores to use self-attention~\cite{vaswani2017attention} based sequential model to capture long-term semantics and use an attention mechanism to make its prediction based on relatively few actions. Besides, there are some other works~\cite{li2020time,hidasi2018recurrent,yu2016dynamic} that introduces specific neural modules for particular recommendation scenarios. For instance, DIN~\cite{zhou2018deep} develops a local activation unit to adaptively learn the user's representation from past behaviors for a specific ad. RUM~\cite{chen2018sequential} introduces a memory-augmented neural network with the insights of collaborative filtering for the recommendation. SDM~\cite{lv2019sdm} integrates a multi-head self-attention module with a gated fusion module to capture both short- and long-term user preferences for the next-item prediction. 

\subsection{Attention Mechanism}
The attention mechanism is initially proposed in computer vision~\cite{burt1988attention} and only becomes popular in recent years. It is first applied to solve the machine translation problem by~\cite{bahdanau2014neural} and later becomes an outbreaking building block as Transformer~\cite{vaswani2017attention}. Recently, BERT leverages Transformer to achieve enormous success in the natural language processing filed for pre-training. It has also been successfully applied in many recommendation applications~\cite{sun2019bert4rec} and is rather useful and efficient in real-world application tasks.

\section{Methodology}\label{sec:prelim}
In this section, we first introduce the problem formulation and then discuss the proposed framework in detail. Finally, we discuss the difference between our framework and existing methods. 

\subsection{Notations and Problem Formulation}
Assume $\{\mathbf{x}^{(u)}\}_{u=1}^N$ be the behavior dataset consists of the interactions between $N$ users and $M$ items. $\mathbf{x}^{(u)}=[x^{(u)}_1, x^{(u)}_2, \cdots, x^{(u)}_n]$ is the ordered sequence of items clicked by user $u$, where $n$ is the number of clicks made by user $u$. Each element $x^{(u)}_t \in \{1,2,\cdots, M\}$ in the sequence is the index of the item being clicked. Note that, due to the strict requirements of latency and performance, industrial recommender systems consist of two stages, the matching stage and ranking stage~\cite{covington2016deep}. The matching stage aims to retrieve top-$N$ candidate items from a large volume of item pool, while the ranking stage targets to sort the candidate items by more precise scores. We focus on improving the effectiveness of the matching stage, where the task is to retrieve high-quality candidate items that the user might be clicked with based on the observed sequence $\mathbf{x}^{(u)}$.

\subsection{Sparse-Interest Framework}
As the item pools of real-world recommender systems often consist of millions or even billions of items, the matching stage is crucial in modern recommender systems. Specifically, a deep sequential model in the matching stage typically has a sequence encoder $\phi_{\theta}(\cdot)$ and an item embedding table $\mathbf{H}\in\mathbb{R}^{M\times D}$, where $\theta$ is the set that contains all the trainable parameters including $\mathbf{H}$. The encoder takes the user's historical behavior sequence $\mathbf{x}^{(u)}$ as input and outputs the representation of the sequence $\phi_{\theta}(\mathbf{x}^{(u)})$, which can be viewed as the representation of the user's intention. The user's intention embedding is then used as a query to generate his/her candidate items from the item pool via a fast K nearest neighbor algorithm (i.e., faiss~\cite{johnson2019billion}). Most encoders $\phi_{\theta}(\cdot)$ in the literature output a single $ D $-dimensional embedding vector, while there are also models that output $ K $ $D$-dimensional embedding vectors to preserve the user's intentions under $K$ latent categories. We mainly focus on the latter direction and target to capture a user's diverse intentions accurately.

The state-of-art sequence encoders for capturing a user's multiple intentions can be summarized into two categories. The first type of methods resort to powerful sequential encoders to \textit{implicitly} extract the user's multiple intentions, such as models based on multi-head self-attention (aka the Transformer~\cite{vaswani2017attention}). The other type of methods rely on the latent prototype to \textit{explicitly} capture a user's multiple intentions. In general, the former approach may limit its ability to capture multiple intentions due to the mixed nature of intention detection and embedding in practice. For example, the empirical results show that the multiple vector representations learned by Transformer do not seem to have a clear advantage over the single-head implementation~\cite{kang2018self} for recommendation.
In contrast, the later can effectively extract a user's diverse interests with the help of concept identified via clustering as empirically proved in~\cite{ma2019learning,liu2019single}. However, these methods scale poor because they require each user has an intention embedding under every concept, which easily scales up to thousands in industrial applications. For instance, millions or even billions of items belong to more than 10 thousand expert-labeled leaf categories~\cite{li2019multi} in the e-commerce platform of Tmall in China. With a large pool of interest concepts in real systems, a scalable multi-interest extraction module is needed. 

Therefore, we propose a sparse-interest network here, which offers the ability to adaptively activate a subset of concepts from the large concept pool for a user. The input of our model is the user's behavior sequence $\mathbf{x}^{(u)}$, which is then fed into an embedding layer and transformed into item embedding matrix $\mathbf{X}^u\in\mathbb{R}^{n\times D}$. Let $\mathbf{C}\in\mathbb{R}^{L\times D}$ denotes the overall conceptual prototype matrix, and $\mathbf{C}^u\in\mathbb{R}^{K\times D}$ indicates the activated prototypical embedding matrix on $K$ latent concepts for user $u$. $L$ is the total number of concepts. 

\subsubsection{Concept activation} Our sparse-interest layer starts by inferring the interested conceptual prototypes $\mathbf{C}^u$ for each user $u$. Given $\mathbf{X}^u\in\mathbb{R}^{n\times D}$, the self-attentive method~\cite{lin2017structured} is first applied to aggregate the input sequence selectively. 
\begin{equation}
\begin{aligned}
\mathbf{a} &= \text{softmax}( \text{tanh}(\mathbf{X}^u\mathbf{W}_1)\mathbf{W}_2),
\end{aligned}
 \label{eq1}
\end{equation}
where $\mathbf{W}_1\in\mathbb{R}^{D\times D}$ and $\mathbf{W}_2\in\mathbb{R}^{D}$ are trainable parameters. The vector $\mathbf{a}\in\mathbb{R}^{n}$ is the attention weight vector of user behaviors. When we sum up the embeddings of input sequence according to the attention weight, we can obtain a virtual concept vector $\mathbf{z}_u=(\mathbf{a}^\top\mathbf{X}^u)^\top$ for the user. $\mathbf{z}_u\in\mathbb{R}^{D}$ reflects the user's general intentions and could be used to activate the interested conceptual prototypes as:
\begin{equation}
\begin{aligned}
\mathbf{s}^u&=\left <\mathbf{C} ,\mathbf{z}_u\right>,\\
\text{idx} &= \text{rank}(\mathbf{s}^u, K),\\
\mathbf{C}^u &= \mathbf{C}(\text{idx}, :) \odot (\text{Sigmoid}(\mathbf{s}^u(\text{idx}, :)\mathbf{1}^T)),
\end{aligned}
 \label{eq2}
\end{equation}
where $\text{rank}(\mathbf{s}^u, K)$ is the top-K ranking operator, which returns the indices of the $K$-largest values in $\mathbf{s}^u$. The index returned by $\text{rank}(\mathbf{s}^u, K)$ contains the indices of prototypes selected for user $u$. $\mathbf{C}(\text{idx}, :)$ performs the row extraction to form the the sub-prototype matrix, while $\mathbf{s}(\text{idx}, :)$ extracts values in $\mathbf{s}^u$ with indices idx. $\mathbf{1}\in\mathbb{R}^K$ is a vector with all elements being 1. $\odot$ represents Hadamard product and $\left <\cdot ,\cdot\right>$ is inner product.
$\mathbf{C}^u\in\mathbb{R}^{K\times D}$ is the final activated $K$ latent concept embedding matrix for user $u$. Equation~\ref{eq2} is a top-$K$ selection trick that enables discrete selection operation differentiable, prior work~\cite{gao2019graph} has found that it is very effective in approximating top-$K$ selection problem. 

\subsubsection{Intention assignment} After inferring the current conceptual prototypes $\mathbf{C}^u$, we can estimate the user intention related with each item in his/her behavior sequence according to their distance to the prototypes.
\begin{equation}
\begin{aligned}
P_{k|t}= \frac{\exp{(\text{LayerNorm}_1(\mathbf{X}^u_t\mathbf{W}_3)\cdot \text{LayerNorm}_2(\mathbf{C}^u_k))}}{\sum_{k'=1}^K \exp{(\text{LayerNorm}_1(\mathbf{X}^u_t\mathbf{W}_3)\cdot \text{LayerNorm}_2(\mathbf{C}^u_{k'}))}},
\end{aligned}
 \label{eq3}
\end{equation}
where $P_{k|t}$ measures how likely the primary intention at position $t$ is related with the $k^{th}$ latent concept. $\mathbf{C}^u_k\in\mathbb{R}^D$ is the embedding of the $k^{th}$ activated conceptual prototype of user $u$. $\mathbf{W}_3\in\mathbb{R}^{D\times D}$ is the trainable weight matrix. $\text{LayerNorm}_l(\cdot)$ represents a layer normalization layer. Note that we are using cosine similarity instead of the inner product here, due to the normalization. This choice is motivated by the fact that cosine is much less vulnerable than dot product when it comes to model collapse~\cite{ma2019learning}, e.g., the degeneration situation where the model is ignoring most prototypes. 

\subsubsection{Attention weighting} In addition to the attention weight $P_{k|t}$ calculated from the conceptual perspective, we also consider another attention weight $P_{t|k}$ to estimate how likely the item at position $t$ is essential for predicting the user's next intentions. 
\begin{equation}
\begin{aligned}
&P_{t|k} = \mathbf{a}^k_t,\\
\mathbf{a}^k &= \text{softmax}( \text{tanh}(\mathbf{X}^u\mathbf{W}_{k,1})\mathbf{W}_{k,2})^T,
\end{aligned}
 \label{eq4}
\end{equation}
$\mathbf{a}^k\in\mathbb{R}^n$ is the attention vector for all positions. The superscript $k$ represents it's the attention layer for the $k^{th}$ activated intention. Similar to Equation~\ref{eq1}, the above equation is another self-attentive layer. The primary difference lies in that we try to make use of the order of user sequences here and add extra trainable positional embeddings~\cite{vaswani2017attention} to the input embeddings. The dimensionality of positional embeddings is the same as that of the item embeddings so that they can be directly summed. 

\subsubsection{Interest embedding generation} We can now generate multiple interest embedding vectors from a user's behavior sequence $\mathbf{X}^u$ according to $P_{k|t}$ and $P_{t|k}$. Specifically, the $k^{th}$ output of our sparse-interest encoder $\phi_{\theta}^k(\mathbf{x}^{(u)})\in\mathbb{R}^D$ is computed as follows:
\begin{equation}
\begin{aligned}
\phi_{\theta}^k(\mathbf{x}^{(u)}) = \text{LayerNorm}_3(\sum_{t=1}^n P_{k|t}\cdot P_{t|k}\cdot \mathbf{X}^u_t).
\end{aligned}
 \label{eq5}
\end{equation}
Till now, we have introduced the whole process of the sparse-interest network. Given a user's behavior sequence, we first activate his/her preferred conceptual prototypes from the concept pool. The intention assignment is then performed to estimate the user intention related with each item in the input sequence. After that, the self-attentive layer is applied to calculate all items' attention weights for next-item prediction. Finally, the user's multiple interest embeddings are generated through a weighted sum, according to Equation~\ref{eq5}.  

\subsection{Interest Aggregation Module}
After the sparse-interest extraction module, we obtain multiple interest embeddings for each user. A natural follow-up question is how to leverage various interest for practical inference. 
An intuitive solution is to use the next predicted item as a target label to select different interest embeddings for training as in MIND~\cite{li2019multi}. Despite its simplicity, the main drawback is that there are no target labels during inference, which leads to a gap between training and testing and may result in performance degeneration. 

To address this issue, we propose an adaptive interest aggregation module based on active prediction. The motivation here is that it is easier to predict a user's temporal preference-based next intentions instead of finding the ideal labels. Specifically, based on the intention assignment score $P_{k|t}$ computed in Equation~\ref{eq3}, we can obtain an intention distribution matrix, denoted by $\mathbf{P}^u\in\mathbb{R}^{n\times K}$, for all items in the behavior sequence. Then, the input behavior sequence $\mathbf{x}^u$ can be reformulated
from the intention perspective denoted by $\widehat{\mathbf{X}^u}=\mathbf{P}^u\mathbf{C}^u$, where $\widehat{\mathbf{X}^u}\in\mathbb{R}^{n\times D}$ is viewed as the intention sequence of user $u$. With $\widehat{\mathbf{X}^u}$, the user's next intention $\mathbf{C}^u_{apt}$ is adaptively computed as  
\begin{equation}
\begin{aligned}
\mathbf{C}^u_{apt} &= \text{LayerNorm}_4\left( (\text{softmax}( \text{tanh}(\widehat{\mathbf{X}^u}\mathbf{W}_3)\mathbf{W}_4))^\top \widehat{\mathbf{X}^u} \right)^\top,
\end{aligned}
 \label{eq6}
\end{equation}
where $\mathbf{C}^u_{apt}\in\mathbb{R}^D$ is the predicted intention of user $u$ for next item. $\mathbf{W}_3\in\mathbb{R}^{D\times D}$ and $\mathbf{W}_4\in\mathbb{R}^D$ are trainable parameters. Given $\mathbf{C}^u_{apt}$ and multiple interest embeddings $\{\phi_{\theta}^k(\mathbf{x}^{(u)})\}_{k=1}^K$, the aggregation weights of different interests are calculated as
\begin{equation}
\begin{aligned}
e^u_k &= \frac{\exp((\mathbf{C}^u_{apt})^\top\phi_{\theta}^k(\mathbf{x}^{(u)})/\tau)}{\sum_{k'=1}^K \exp((\mathbf{C}^u_{apt})^\top\phi_{\theta}^{k'}(\mathbf{x}^{(u)})/\tau)}.
\end{aligned}
 \label{eq7}
\end{equation}
Where $e^u=[e^u_1,e^u_2,\cdots,e^u_K]^T\in\mathbb{R}^{K}$ is the attention vector for diverse interests. $\tau$ is a temperature parameter to tune. When $\tau$ is large ($\tau \rightarrow \infty$), $e^u$ approximates a uniformly distributed vector. When $\tau$ is small ($\tau \rightarrow 0^+$), $e^u$ approximates a one-hot vector. In experiments, we use $\tau=0.1$ to enforce the aggregator select the most preferred intention for inference. The final user representation $\mathbf{v}^u\in\mathbb{R}^D$ is computed as
\begin{equation}
\begin{aligned}
\mathbf{v}^u=\sum_{k=1}^K e^u_k \cdot \phi_{\theta}^k(\mathbf{x}^{(u)}.
\end{aligned}
 \label{eq8}
\end{equation}
\subsection{Model Optimization}
We follow the common practice~\cite{kang2018self,li2019multi} to train our model by recovering the next click $x^{(u)}_t$ based on the truncated sequence prior to the click, i.e., $[x^{(u)}_1,x^{(u)}_1,\cdots,x^{(u)}_{t-1}]$. Given a training sample $(u,t)$ with the user embedding vector $\mathbf{v}^u$ and item embedding $\mathbf{H}_t$, we aim to minimize the following negative log-likelihood
\begin{equation}
\begin{aligned}
\mathcal{L}_{like} & = -\sum_{u}\sum_t\log P(x^{(u)}_t|x^{(u)}_1,x^{(u)}_2,\cdots,x^{(u)}_{t-1})  \\
& = -\sum_{u}\sum_t \log \frac{\exp(\mathbf{H}_t^\top\mathbf{v}^u)}{\sum_{j\in\{1,2,\cdots, M\}}\exp(\mathbf{H}_{j}^\top\mathbf{v}^u))}.
  \label{eq9}
 \end{aligned}
\end{equation}
Equation~\eqref{eq9} is usually intractable in practice, because the sum operation of the denominator is computationally prohibitive. We, therefore, leverage a  Sampled Softmax technique~\cite{covington2016deep,jean2014using} to train our model. 
Besides, we also introduce a covariance regularizer following~\cite{cogswell2015reducing} to enforce the learned conceptual prototypes orthogonally. Specifically, denote $\mathbf{M}=\frac{1}{D}(\mathbf{C}-\overline{C})(\mathbf{C}-\overline{C})^\top$ as the covariance matrix of prototype embeddings, where $\overline{\mathbf{C}}$ is the mean matrix of $\mathbf{C}$. The regularization loss $\mathcal{L}_c$ to regularize the covariance is
\begin{equation}
\begin{aligned}
\mathcal{L}_c & = \frac{1}{2}(||\mathbf{M}||_F^2-||\text{diag}(\mathbf{M})||^2_F).
  \label{eq10}
 \end{aligned}
\end{equation}
Where $||\cdot||_F$ is the Frobenius norm matrix. Combine the two losses above, the final loss function of our model is
\begin{equation}
\begin{aligned}
\mathcal{L} & = \mathcal{L}_{like} + \lambda \mathcal{L}_c,
  \label{eq11}
 \end{aligned}
\end{equation}
where $\lambda$ is the trade-off parameter to balance the two losses. 

\subsection{Connections with Existing Models}
We compare our model and existing methods that focus on extracting user's multiple interest embeddings in the matching stage of recommendation. We roughly divided them into two categories and analyzed the difference below. 

\noindent \textbf{Implicit approach}. This type of method relies on powerful neural networks to implicitly cluster historical behaviors and extract diverse interests. For example, MIND~\cite{li2019multi} utilizes Capsule network~\cite{sabour2017dynamic} to adaptively aggregate user's behaviors into interest embedding vectors. SASRec~\cite{kang2018self} adopts the multi-head self-attention mechanism~\cite{vaswani2017attention} to output multiple representation for a user. Compared with these methods, our model belongs to an explicit approach that explicitly detects intentions from the user's behavior sequence based on latent conceptual prototypes. 

\noindent\textbf{Explicit approach}. Methods that belong to this type maintain a set of conceptual prototypes to explicitly determine the intentions of items in the user's behavior sequence. MCPRN~\cite{wang2019modeling} is a recent representative work for extracting multiple interests from the session for the next-item recommendation. DisenRec~\cite{ma2019learning} utilizes latent prototypes to help learn disentangled representations for recommendation. Compared with them, we also follow the explicit approach, but our model scales to a large-scale dataset. Specifically, they require the number of diverse interest embeddings equals to the number of 
conceptual prototypes. However, the number of latent concepts depends on applications and can be easily scaled up to hundreds or even thousands in industrial recommender systems, which hinders their application in practice. In contrast, our sparse-interest network offers the ability to infer a sparse set of preferred intentions from the large concept pool automatically.

\section{Experiments}\label{sec:experiment}
In this section, we conduct experiments over three benchmark datasets and one billion-scale industrial data to validate the proposed approach. Specifically, we try to answer the following  questions: 
\begin{table*}[h!]
  \caption{Recommendation performance on public datasets. The best results are highlighted with bold fold. All the numbers in the table are percentage numbers with '\%' omitted.  }
  {
    \begin{tabular}{c|cccc|cccc|cccc}
    \toprule
    \multirow{3}*{} &\multicolumn{4}{c}{\textbf{MovieLens}} &\multicolumn{4}{c}{\textbf{Amazon}} &\multicolumn{4}{c}{\textbf{Taobao}} \\
    
    &\multicolumn{2}{c}{Metrics@10}
    &\multicolumn{2}{c}{Metrics@50}
    &\multicolumn{2}{c}{Metrics@50}
    &\multicolumn{2}{c}{Metrics@100}
    &\multicolumn{2}{c}{Metrics@50}
    &\multicolumn{2}{c}{Metrics@100}\\
    
   \hline
    & HR &NDCG &HR &NDCG &HR &NDCG &HR &NDCG &HR &NDCG &HR &NDCG\\
    \hline
    \textbf{GRU4Rec} &14.61 &5.66 &41.61 &10.66 &1.70 &0.51 &2.74 &0.67 & 9.41 &3.60 &12.43 &4.08\\
     \textbf{Caser} &15.44 &6.13 &43.64 &11.53 &2.60 &0.81 &3.96 &1.03 & 10.71 &4.96 &13.50 &5.68\\
    \textbf{SASRec} &\textbf{17.34} &\textbf{7.84} &\textbf{46.01} &\textbf{13.53} &3.17 &1.01 &4.43 &1.28 &13.36 &5.64 &15.73 &6.38\\
    \textbf{MIND} &15.62 &6.58 &43.98 &12.30 &3.85 &1.29 &5.35 &1.56 & 15.35 &8.35 &17.49 &8.72\\
    \textbf{MCPRN} &15.82 &6.77 &44.21 &12.83 &3.42 &1.18 &5.22 &1.47 &14.32 &7.34 &16.43 &7.67\\
    \hline
    \textbf{SINE} &16.34 &7.06 &\textbf{45.79} &\textbf{13.50} &\textbf{4.57} &\textbf{1.61} &\textbf{6.26} &\textbf{1.88} & \textbf{17.69} &\textbf{10.41} &\textbf{20.64}  &\textbf{10.89}\\
  \bottomrule
\end{tabular}}
\label{table2}
\end{table*}

\begin{itemize}
    \item How effective is the proposed method compared to other state-of-the-art baselines? \textbf{Q1}
    \item What are the effects of the different modules, sparse-interest module, and interest aggregation module through ablation studies? \textbf{Q2}
     \item How sensitive are the hyper-parameter settings, including the preferred $ K $ intentions and the corresponding $ L $ conceptual prototypes? \textbf{Q3}  
\end{itemize}

\begin{table}[htbp]
  \caption{Statistics of the datasets.}
  \begin{tabular}{c| c |c |c}
   \toprule
    Dataset &\# users &\# items &\# interactions\\
     \hline
     MovieLens &6,040 &3,952 &1,000,209\\
    Amazon &8,026,324 &2,330,066 &22,507,155\\
    Taobao & 987,994 &  4,162,024 &100,150,807\\
    ULarge &106,527,123  &25,000,000 &4,000,000,000\\
 \bottomrule
\end{tabular}
\label{table1}
\vspace{-0.1in}
\end{table}
\subsection{Experimental Setup}
In this section, we elaborate on the dataset description, evaluation metrics, and comparing methods in our experiments.

\noindent\textbf{Datasets}. We conduct experiments on three benchmark datasets and one billion-scale industrial data. The statistics of the datasets are shown in Table~\ref{table1}. 
\begin{itemize}
\item \textbf{MovieLens}~\footnote{https://grouplens.org/datasets/movielens/1m/} collects user's rating score for movies. In experiments, we follow~\cite{he2016fast} to preprocess the dataset. 

\item \textbf{Amazon}~\footnote{http://jmcauley.ucsd.edu/data/amazon/} consists of product views from Amazon. In experiments, we use the rating only version of Book category behaviors. Note that this version is more challenging than the 5-core version used in~\cite{li2019multi}, due to its large volume and sparsity. 

\item \textbf{Taobao}~\footnote{https://tianchi.aliyun.com/dataset/dataDetail?dataId=649} collects user behaviors from Taobao's recommender system. In experiments, we only use the click behaviors. 

\item \textbf{ULarge} consists of the clicked behaviors collected from the daily logs of an Alibaba company from March 29 to April 4, 2020. 
\end{itemize}

For all datasets, we follow~\cite{kang2018self} to split the datasets into training/validation/testing sets. Specifically, we split the historical sequence for each user into three parts: (1) the most recent action for testing, (2) the second most recent action for validation, and (3) all remaining actions for training. Note that during testing, the input sequences contain training actions and the validation actions.

\noindent\textbf{Competitors}. We compare our proposed model SINE with the following state-of-the-art sequential recommendation baselines.
\begin{itemize}
\item \textbf{Single embedding models}: GRU4Rec~\cite{hidasi2015session} is a pioneering work that employs GRU to model user behavior sequences. 
Caser~\cite{tang2018personalized} is a recent CNN-based sequential recommendation benchmark.

\item \textbf{Multi-embedding models}: MIND~\cite{li2019multi} and SASRec~\cite{kang2018self} are recently proposed multi-interest methods based on capsule network~\cite{sabour2017dynamic} and multi-head self-attention~\cite{vaswani2017attention}. MCPRN is another state-of-the-art multi-interest framework based on latent conceptual prototypes.    
\end{itemize}

\noindent\textbf{Parameter Configuration}. For a fair comparison, all methods are implemented in Tensorflow and optimized with Adam optimizer with a mini-batch size of 128. The learning rate is fixed as 0.001. We tuned the parameters of comparing methods according to values suggested in original papers and set the embedding size $ D $ as 128 and the number of negative samples as 5 and 10 for MovieLens and other datasets. For our method, it has three crucial hyper-parameters: the trade-off parameter $ \lambda $, the number of intentions $ K $, and latent prototypes $ L $. We search $K$ from $\{4, 8, 12, 16\}$, $L$ from $\{50, 100, 500, 1000, 2000, 5000\}$ and $\lambda$ from 0 to 1 with step size 0.1. We found our model performs relative stable when $\lambda$ is around 0.5 and set $\lambda=0.5$. The configuration of the other two parameters for four datasets are reported in Table~\ref{table3}. 
\begin{table}[h]
  \caption{The optimal setting of our hyper-parameters for our model. Other parameters like dimension $D$, sequence length $n$ and $\lambda$ are set as 128, 20 and 0.5, respectively.}
  \begin{tabular}{c|c|c}
    \toprule
     & \# intentions $K$ & \# concepts $L$ \\
    \hline
    MovieLens &4 &50 \\
    Amazon &4 &500 \\
    Taobao &8 &1000 \\
    ULarge &8 &5000 \\
  \bottomrule
\end{tabular}
\label{table3}
\end{table}

\noindent\textbf{Evaluation Metrics}.
For each user in the test set, we treat all the items that the user has not interacted with as negative items. We use two commonly used evaluation criteria~\cite{he2017neural}: \textit{hit rate} (HR) and \textit{normalized discounted cumulative gain} (NDCG) to evaluate the performance of our model. Besides, we also leverage the widely used \textit{Normalized Mutual Information} (NMI)~\cite{mcdaid2011normalized} to quantitative analysis of the effectiveness of the learned conceptual prototypes of our model in clustering items.

\begin{figure*}[htbp]
  \centering 
  \includegraphics[width=18.0 cm,height=7 cm]{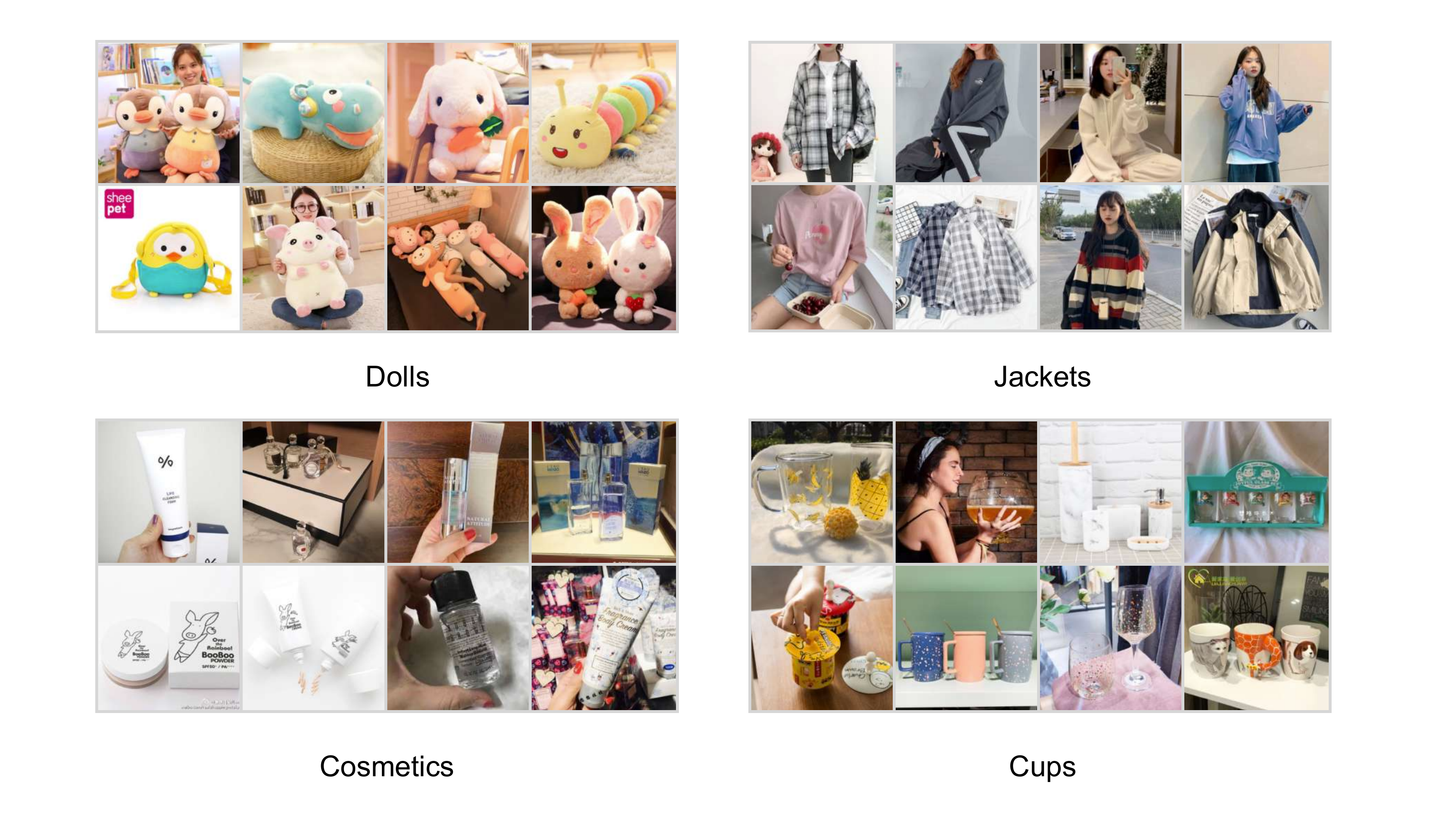}
  \caption{Concept visualization. We draw four concepts "dolls", "jackets", "cosmetics" and "cups" with the top-8 closest items.}
  \label{figure4}
\end{figure*}
\subsection{Comparisons with SOTA (Q1)}
Table~\ref{table2} summarizes the performance of SINE as well as baselines on three benchmark datasets. Clearly, SINE achieves comparable performance to all of the baselines on all the evaluation criteria in general. Caser obtains the best performance over other models (GRU4Rec) that only single output embedding for each user. It can be observed that employing multiple embedding vectors (SASRec, MIND, MCPRN, SINE) for a user perform generally better than single embedding based methods (Caser and GRU4Rec). Therefore, exploring multiple user-embedding vectors has proved to be an effective way of modeling users' diverse interests and boosting sequential recommendation accuracy. Moreover, we can observe that the improvement introduced by capturing user's various intentions is more significant for Taobao and Amazon datasets. The users of Taobao and Amazon tend to exhibit more diverse interests in online shopping than rating movies. The improvement of MIND over SASRec shows that dynamic routing serves as a better multi-interest extractor than multi-head self-attention. An interesting observation is that MIND beats MCPRN on Amazon and Taobao while losses on MovieLens. It is mainly because MCPRN only supports cluster all items into a small set of prototypes, which is difficult well to cluster millions of items on Amazaon and Taobao. Considering the MIND and SINE results, SINE consistently outperforms MIND on three datasets over all evaluation metrics. This can be attributed to two points: 1) The sparse-interest extractor layer explicitly utilizes a large set of conceptual prototypes to cluster items and automatically infer a subset of preferred intentions for interest embeddings generation, which achieves a more precise representation of a user. 2) Interest aggregation module actively predicts the user's current intention to directly attend over multiple user embedding vectors, enabling modeling multi-interests for top-N recommendation. 

\begin{figure}
\centering
\subfloat{{\includegraphics[width=4.2cm]{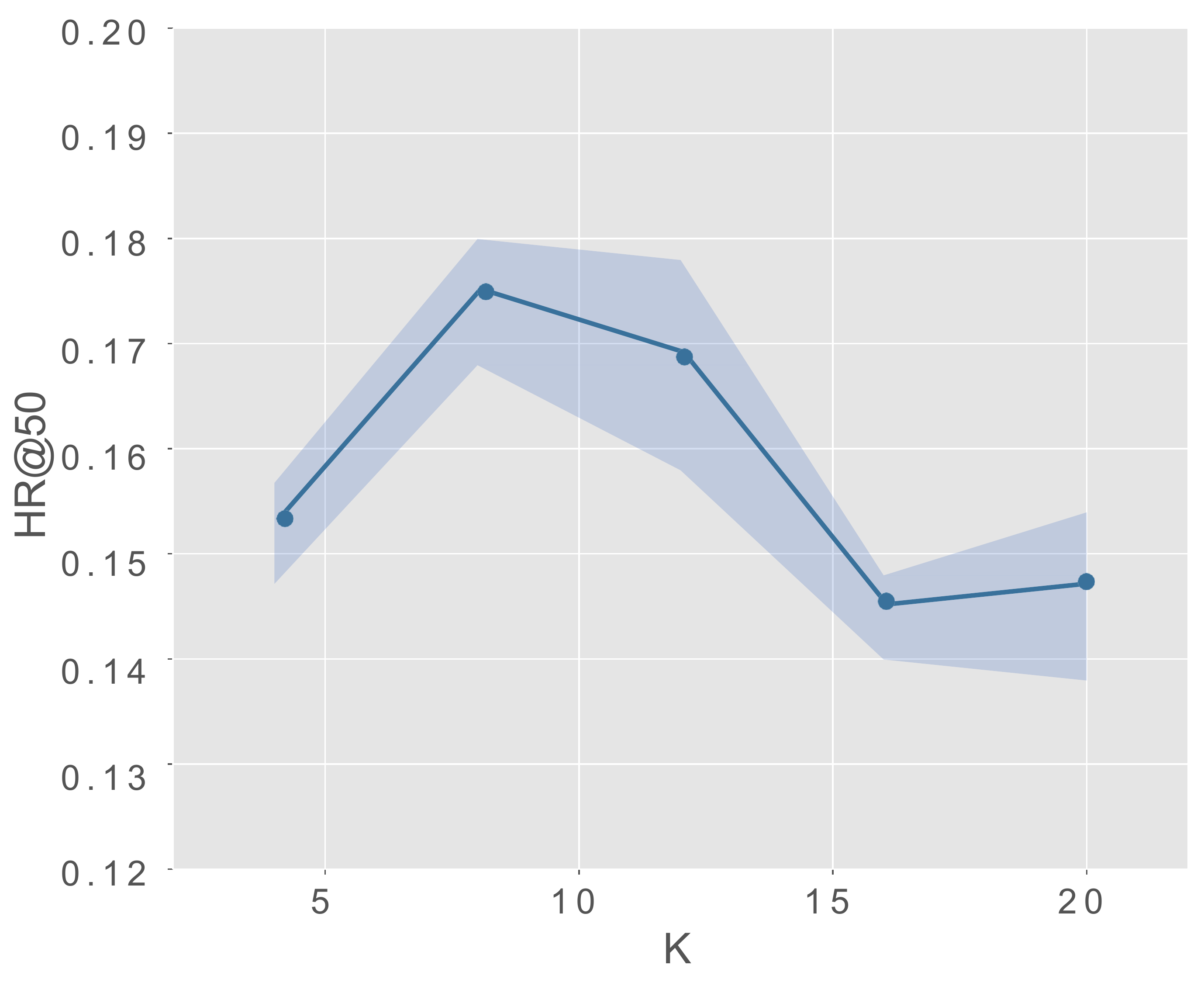}}}%
\subfloat{{\includegraphics[width=4.2cm]{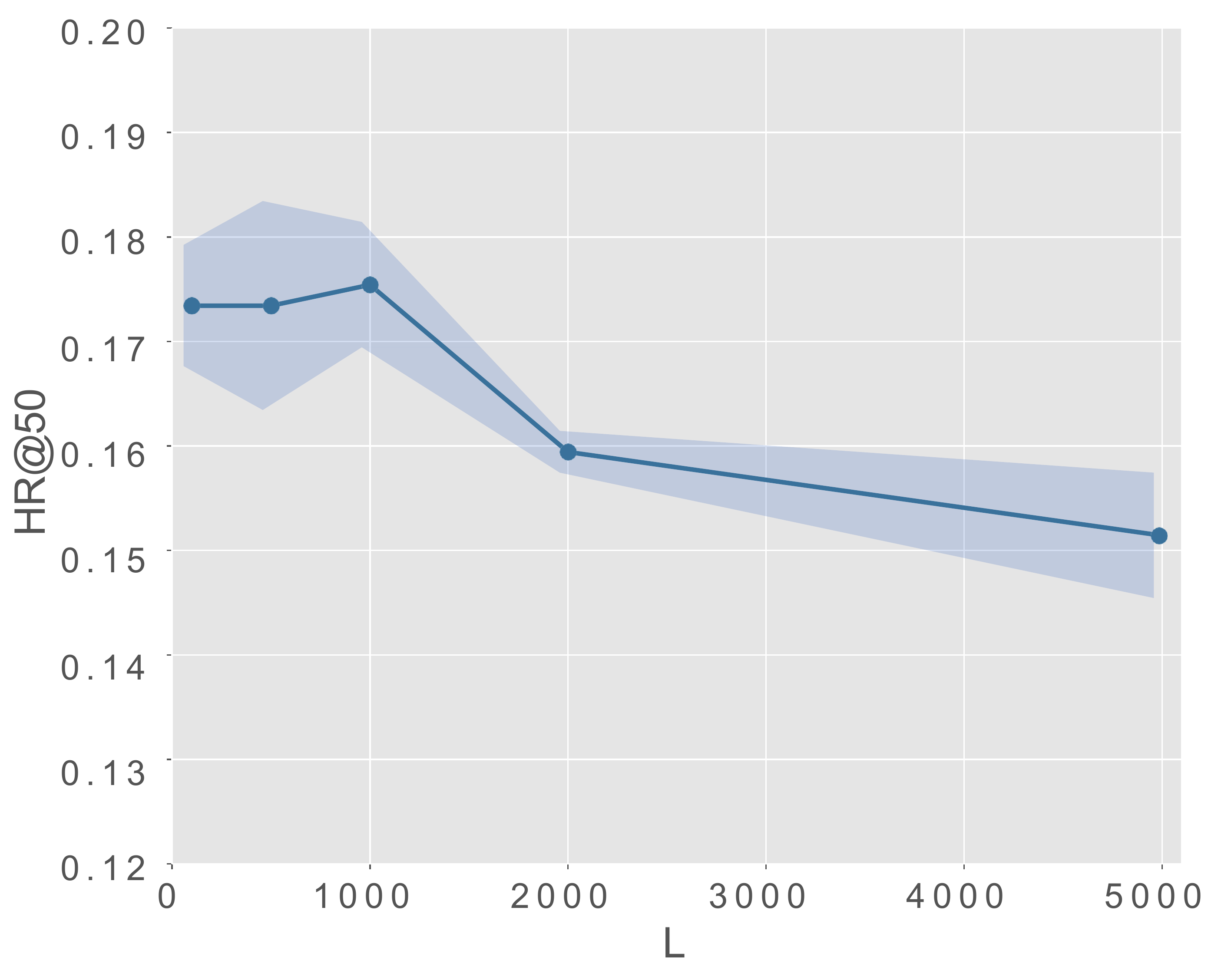}}}%
\caption{Sensitivity of SINE towards $K$ and $L$ on Taobao.}
\label{fig2}
\end{figure}

\noindent\textbf{Parameter Sensitivity (Q3)}. We also investigate the sensitivity of the number of intentions $K$ and conceptual prototypes $L$. Figure~\ref{fig2} reports the performance of our model in terms of HR. In particular, we randomly select 1 million users for inference, and the average result of 10 runs is reported. Results hold the same for other datasets, and we omit the figure here for more space. From the figure, we can observe that SINE obtains the best performance when $K=8$ and $L=1000$. Considering that Taobao has around 9000 different categories in total, it verifies that the learned concepts indeed have a strong connection of categories of items, and the concept could be viewed as a virtual category that consists of several categories.  

\begin{table}[htbp]
  \caption{Recommendation performance on industrial dataset ULarge. Improv. row means the improvement of our model compared with the second-best baseline.}
  \begin{tabular}{lcccc}
    \toprule
     &HR@50 &HR@100 &HR@500 \\
    \hline 
    \textbf{Caser} &6.93 &16.75 &36.94 \\
    \textbf{GRU4Rec} &5.46 &14.80 &33.35 \\
    \textbf{SASRec} &8.64 &18.58 &38.82 \\
    \textbf{MCPRN} &7.89  &17.65  &37.66  \\
    \textbf{MIND} &9.13 &19.31 &39.09 \\
    \textbf{SINE} &\textbf{12.24}  &\textbf{21.12}  &\textbf{40.81}  \\
    \hline 
    \textbf{Improv.} &34.06\% &9.37\% &4.40\%\\
  \bottomrule
\end{tabular}
\label{table4}
\end{table}

\subsection{Industrial Results (Q1)}
We further conduct an offline experiment to investigate the effectiveness of our model in extracting user's diverse interests in the industrial dataset. We implemented our model and baselines on the Alibaba company's distributed cloud platform, where every two workers share an NVIDIA Tesla P100GPU with 16GB memory. 

Table~\ref{table4} summarizes the performance in terms of Hit Rate. It is clear that SINE significantly outperforms other baselines by a wide margin. Another interesting observation is that the gap between SINE and the second-best benchmark (MIND) decreases when the number of recalled items increases. This fact indicates that our sparse-interest network helps capture user's diverse interests and ranks the most preferred items on the top recommendation list. 

\noindent\textbf{Case Study}
We also visualize the learned conceptual prototypes of our model. Concretely, for each concept, we leverage its prototypical embedding vector to retrieve the top-8 closest items under their cosine similarity. Figure~\ref{figure4} illustrates four exemplar concepts to show their clustering performance.   
As can be seen, our model successfully groups some semantic-similar items into a latent concept. More importantly, the items in one concept come from different semantic-close leaf categories. For example, the ``cosmetics" concept contains different kinds of skin-nursing products. It indicates that compared to the conventional leaf-category partition, our conceptual prototype is related to the user's high-level intention.  

To confirm this point, we compare the learned concepts with the expert-labeled category hierarchy in Alibaba company, where the number of categories in the first, second, and leaf-level are 178, 7,945, and 14874, respectively. Table~\ref{table5} reports the results in terms of NMI. We can observe that the learned concepts are closest to the second level category, not in the extreme fine-grained granularity (leaf) or the very coarse granularity (first). This result demonstrates that our model can capture the relative high-level semantics for the user's intention modeling. 

\begin{table}[!t]
  \caption{Prototype clustering evaluation compared with the first, second and leaf level category information on ULarge.}
  \begin{tabular}{lccc}
    \toprule
         & Level-1 & Level-2 & Level-leaf \\
    \hline
     NMI & 0.09 & 0.37 & 0.29\\
  \bottomrule
\end{tabular}
\label{table5}
\end{table}

\begin{table}
  \caption{Ablation study of SINE.}
  \begin{tabular}{cccc}
    \toprule
     \textbf{Dataset} &\textbf{Method} &\textbf{HR@50} &\textbf{HR@100}\\
    \midrule
    \multirow{3}*{\textbf{Taobao}}
    &\textbf{SINE-cate} &12.45 &15.33 \\
    &\textbf{SINE-label} &16.22 &18.74\\
    &\textbf{SINE} &\textbf{17.69} &\textbf{20.64}\\
    \midrule
    \multirow{4}*{\textbf{ULarge}}
    &\textbf{SINE-cate} &7.18 &17.46 \\
    &\textbf{SINE-label} &10.09 &20.33\\
    &\textbf{SINE} &\textbf{12.24} &\textbf{21.12}\\
  \bottomrule
\end{tabular}
\label{table6}
\end{table}

\subsection{Ablation Study (Q2)}
We introduce two variants (SINE-cate and SINE-label) to validate the effectiveness of the learned new prototypes and the interest aggregation module. Specifically, SINE-cate is obtained by using the category attributes as prototypes, while SINE-label is obtained by adopting label-aware attention in~\cite{li2019multi} for training. We only conduct experiments on Taobao and ULarge, since other datasets do not have category attributes. Taobao and ULarge have 9439 and 14874 distinct categories, respectively. Note that, similar to MIND~\cite{li2019multi}, SINE-label first independently retrieves $K\cdot$ N candidate items based on $K$ embedding vectors and then outputs the final top-N recommendation list by sorting $K\cdot$ N items. Table~\ref{table5} reports the results in terms of HR. Obviously, SINE significantly outperforms the other two variants in two datasets. The substantial difference between SINE-cate and SINE shows that the learned concepts are better to cluster items than the original items' categories. It verifies our motivation to cluster items in our model jointly. The improvement of SINE over SINE-label validates that our interest attention module is useful to model multiple interests for next-item recommendation.

\section{Conclusions}\label{sec:experiment}
In this paper, we propose a novel sparse-interest embedding framework for the sequential recommendation. Our model can adaptively activate multiple intentions from a large pool of conceptual prototypes to generate multiple interest embeddings for a user. It also develops an interest aggregation module to capture multi-interests to obtain the overall top-N items actively. Empirical results demonstrate that our model performs better than state-of-the-art baselines on challenging datasets. Results on the billion-scale industrial dataset further confirm our model's effectiveness in terms of recommendation accuracy and producing reasonable item clusters. We plan to leverage lifelong learning to capture users' long-term interests for a more accurate recommendation.  

\normalem
\bibliographystyle{ACM-Reference-Format}
\balance
\bibliography{sample-base}

\end{document}